%% file: SusEng_Main.tex
\documentclass[10pt, conference, letterpaper]{IEEEtran}
\IEEEoverridecommandlockouts

\usepackage{amsfonts, amsmath,amssymb,amsthm}
\usepackage{algorithm, algorithmic}
\usepackage{array}
\usepackage{cases}
\usepackage{changepage}
\usepackage[compress]{cite}
\usepackage{etoolbox}
\usepackage{graphicx}
\usepackage{multirow}
\usepackage{physics} 	
\usepackage{subfigure}
\usepackage{setspace}
\usepackage{url}  
\usepackage{lipsum}
\usepackage[table,xcdraw]{xcolor}	
\usepackage{stmaryrd} 

\def\ps@headings{%
\def\@oddhead{\mbox{}\scriptsize\rightmark \hfil \thepage}%
\def\@evenhead{\scriptsize\thepage \hfil \leftmark\mbox{}}%
\def\@oddfoot{}%
\def\@evenfoot{}}
\makeatother
\makeatletter
\patchcmd{\@thm}
  {\trivlist}
  {\list{}{\leftmargin=\thm@leftmargin\rightmargin=\thm@rightmargin}}
  {}{}
\patchcmd{\@endtheorem}
  {\endtrivlist}
  {\endlist}
  {}{}
\newlength{\thm@leftmargin}
\newlength{\thm@rightmargin}

\newcommand{\xnewtheorem}[3]{%
  \newenvironment{#3}
    {\thm@leftmargin=#1\relax\thm@rightmargin=#2\relax\begin{#3INNER}}
    {\end{#3INNER}}%
  \newtheorem{#3INNER}%
}

\makeatother

\newtheoremstyle{indentedupright}{3pt}{3pt}{} {}{\bfseries}{.}{.5em}{} 
\newtheoremstyle{indenteditalic}{3pt}{3pt}{\itshape} {}{\bfseries}{.}{.5em}{} 

\theoremstyle{indenteditalic}
\xnewtheorem{0pt}{0pt}{oldpro}{Conventional Property}
\xnewtheorem{0pt}{0pt}{pro}{Property}
\xnewtheorem{0pt}{0pt}{newpro}{New Property}
\xnewtheorem{0pt}{0pt}{obser}{Observation}

\renewcommand{\baselinestretch}{1}

\begin{document}
\title{\LARGE Practical Issues of Energy Harvesting and Data Transmissions in Sustainable IoT}
\author{\IEEEauthorblockN{Yu Luo\IEEEauthorrefmark{1}, Lina Pu\IEEEauthorrefmark{2}}
\IEEEauthorblockA{\IEEEauthorrefmark{1}ECE Department, Mississippi State University, Mississippi State, MS, 39762.\\
\IEEEauthorrefmark{2}School of CSCE, University of Southern Mississippi, Hattiesburg, MS 39406.}
}

\maketitle
\thispagestyle{plain} 
\pagestyle{plain}       
\input{Abstract.tex}

\input{Introduction.tex}

\input{RelatedWork.tex}

\input{Environments.tex}

\input{EngHarModel.tex}

\input{EngSingle.tex}

\input{EngContinuous.tex}

\input{EngMod.tex}

\input{Conclusion.tex}

\input{Appendix.tex}

\bibliographystyle{IEEEtran}
\bibliography{SusEng_Main}

\end{document}

%% file: Abstract.tex
\begin{abstract}
\label{sec:abstract}
The sustainable Internet of Things (IoT) is becoming a promising solution for the green living and smart industries. In this article, we investigate the practical issues in the radio energy harvesting and data communication systems through extensive field experiments. A number of important characteristics of energy harvesting circuits and communication modules have been studied, including the non-linear energy consumption of the communication system relative to the transmission power, the wake-up time associated with the payload, and the varying system power during consecutive packet transmissions. In order to improve the efficiency of energy harvest and energy utilization, we propose a new model to accurately describe the energy harvesting process and the power consumption for sustainable IoT devices. Experiments are performed using commercial IoT devices and RF energy harvesters to verify the accuracy of the proposed model. The experiment results show that the new model matches the performance of sustainable IoT devices very well in the real scenario.  

\end{abstract} 

%% file: Introduction.tex
\section{Introduction}
\label{sec:intro}

Due to the ever-increasing demand for green and sustainability, the renewable energy harvesting technology has attracted extensive attention for its potential to self-power a large number of IoT devices. Recently,  many semiconductor modules have been developed by both academia and industry to efficiently capture renewable radio frequency (RF) energy radiated from TV towers, WiFi base stations, and wireless routers~\cite{muncuk2018multiband, kim2014ambient, powercast2016p2110b}. The received energy can be stored in an energy storage component (ESC), like a supercapacitor or a small rechargeable battery, to power the target systems for control, sensing, and wireless communications.


The key to IoT technology is the wireless communication capability, which significantly broadens the application's scope of IoT technology. However, the communication link may be the most power-hungry part of a typical IoT device. In order to meet the communication quality requirements under strict energy constraints, system designers need to know the maximum throughput of each RF powered device because it determines the best performance for IoT applications. However, the throughput estimation requires the information about the exact amount of energy that can be harvested and will be consumed by the communication link, which is a challenge in the real world due to the high dynamic RF environment and complex relationship between the power consumption and the hardware design of the IoT device. 

Modeling the power consumption of communication modules and the energy harvesting process of sustainable IoT devices is an important but challenging task that has not been fully studied in the literature. First of all, the system power consumption increases nonlinearly with the transmission power of a transceiver; how to properly describe their relationship should be carefully considered in a power consumption model. Second, as will be introduced in this article, the energy overhead involved in each data transmission is not a constant but changes with the ESC voltage level and the payload in a packet. Accurately calculating the energy overhead under different protocol configurations and hardware settings is a challenge. Third, due to the environment-dependent energy conversion efficiency of the harvesting circuit and the nonlinear charging characteristic of the ESC, the energy harvesting rate of a sustainable IoT device becomes a complex function of the RF power density and the energy harvesting time. Therefore, it is challenging to accurately estimate the energy harvesting rate in a dynamic RF environment.

To tackle the above challenging problems, we conduct extensive experiments to study the unique features inherited from the hardware and protocols of sustainable IoT system. In the experiment, a Powercast P2110 module~\cite{powercast2016powercast} is used to scavenge renewable radio energy from surrounding environments. The collected energy is then stored in a supercapacitor to power the Microchip ATmega256RFR2 chipset~\cite{microchip2016atmel} that serves as the core of an IoT device for system control, environment sensing, and wireless communication. At the physical (PHY) layer and the medium access control (MAC) layer, we run IEEE standard 802.15.4 with different transmission powers, data rates, and effective payloads to evaluate the accuracy of the proposed model under different settings.

Several new important conclusions can be drawn from the experimental results. First of all, although the intensity of renewable RF energy has significantly random fluctuations over time, the battery charging process is very smooth. This observation provides an IoT device an opportunity to accurately predict the amount of energy that can be received in the future, which is very useful for the design of online power management and data transmission strategies. Second, in many commercial IoT devices, the transmission power can be expressed as a modified sigmoid function of system power consumption. This shows that we can accurately calculate the system power consumption given a transmission power. Third, the wake-up time of IoT devices is not a fixed value, but increases linearly with the data payload. Therefore, reducing the packet size can wake the device from the sleep mode faster. Finally, given a transmission power and a data rate, the total energy consumed by an IoT device on transmitting a packet gradually reduces over time due to the decreasing voltage of the ESC. This should be considered in power management to avoid overestimating energy consumption on continuous packet transmission.

The new model proposed in this paper takes into account the characteristics of the renewable RF energy, harvester circuit, and communication protocol to accurately describe the energy harvesting process and the power consumption of renewable RF energy powered IoT devices. The results obtained in this paper can be used to design more realistic and efficient strategies for power management and transmission scheduling. The proposed model can help develop hardware and communication protocols for sustainable IoT devices to improve system performance in terms of energy harvesting efficiency, throughput, and efficient energy utilization.

The rest of the paper is organized as follows. Section~\!\ref{sec:RelatedWork} presents the related work. We investigate the characteristics of renewable RF energy in Section~\!\ref{sec:Env}. The RF energy harvesting model is introduced in Section~\!\ref{sec:EngHarPro}. Afterward, we analyze the energy consumption of sustainable IoT device on single packet transmission and continuous packet transmission in Section~\!\ref{sec:EngSig} and Section~\!\ref{sec:EngCon}, respectively. In Section~\!\ref{sec:EngMod}, we introduce the  energy consumption model for sustainable IoT devices. Conclusions are presented in Section~\!\ref{sec:conclusion}.

%% file: RelatedWork.tex
\section{Related Work}
\label{sec:RelatedWork}
In recent years, many models have been developed to describe the energy harvesting process and power consumption of renewable radio energy powered IoT devices~\cite{ulukus2015energy}. In these models, it is usually assumed that the received energy are discrete energy packets with random sizes and arrival time~\cite{chen2015provisioning, yang2012optimal}. With this assumption, an efficient data transmission strategy is converted to a segmentation optimization problem to find the appropriate transmission power in an irregular energy tunnel. As introduced in \cite{tutuncuoglu2012optimum}, after considering the ESC capacity constraint and the energy harvesting causality, the profile of the cumulative energy consumption of sustainable IoT devices should be the tightest string in the energy tunnel to maximize throughput. 

In order to make the data transmission strategy realistic, more and more hardware characteristics of the RF energy harvester are considered in recent energy harvesting and power consumption models. According to the results published in \cite{luo2018revisiting} and \cite{gorlatova2013networking}, the amount of harvested energy depends not only on the strength of RF power density but also on the amount of residual energy in the ESC. In \cite{biason2016effects}, it is assumed that an energy harvester has only an incomplete knowledge about the ESC energy level. In this case, the power management can be modeled as a partially observable Markov decision process to maximize long-term throughput. When applying this method, the energy harvester does not measure the energy level of the ESC directly, but rather maintains a probability distribution of the remaining energy in a set of possible states based on the observation of the energy harvested in the past period. In \cite{zhao2019power}, the overhead energy of IoT device and the energy loss of the ESC over time are taken into account to improve the accuracy of the energy consumption model. However, the energy consumption of the circuit in this work is simply treated as a constant rather than an environment-dependent variable, which in some cases may not match the real scene.

In \cite{chaour2017enhanced}, the impacts of the diode nonlinearity and parasitic effects in the energy harvesting circuit on the radio energy harvesting efficiency is investigated. The result reveals that the intensity of the incident radio waves can significantly affect the power conversion efficiency of RF harvesters. In addition, energy leakage caused by the off current in the circuit and the self-discharge characteristic of the ESC is another factor may affect system performance. The results obtained in \cite{devillers2012general} show that the impact of  energy leakage on a transmission strategy is equivalent to adding a constant operation power on the circuit; accordingly, the width of the energy tunnel becomes narrow over time. To correctly estimation energy consumption of an IoT device, the overhead energy generated by the microprocessor and associated peripherals should be taken into account carefully. As analyzed in \cite {xu2014throughput} and \cite {orhan2014energy}, after considering the energy overhead, the strategy of continuous data transmission becomes inefficient, and an on-off transmission strategy is proposed for throughput optimization.

From the literature review, we realized that most of the existing works are based on assumptions rather than real data to simulate the energy harvesting and energy consumption of RF energy powered IoT devices. However, according to the experimental results, we realized that some assumptions may be inaccurate. Therefore, a new model is needed, which is the focus of this paper.

%% file: Environments.tex
\section{Overview of Renewable RF Energy}
\label{sec:Env}
In this section, we study the characteristics of renewable radio energy. Based on measurements, several important features of outdoor RF energy will be summarized to help us model the renewable RF energy harvesting process in the real wold. 

\begin{figure}[htb]
\centerline{\includegraphics[width=8.7cm]{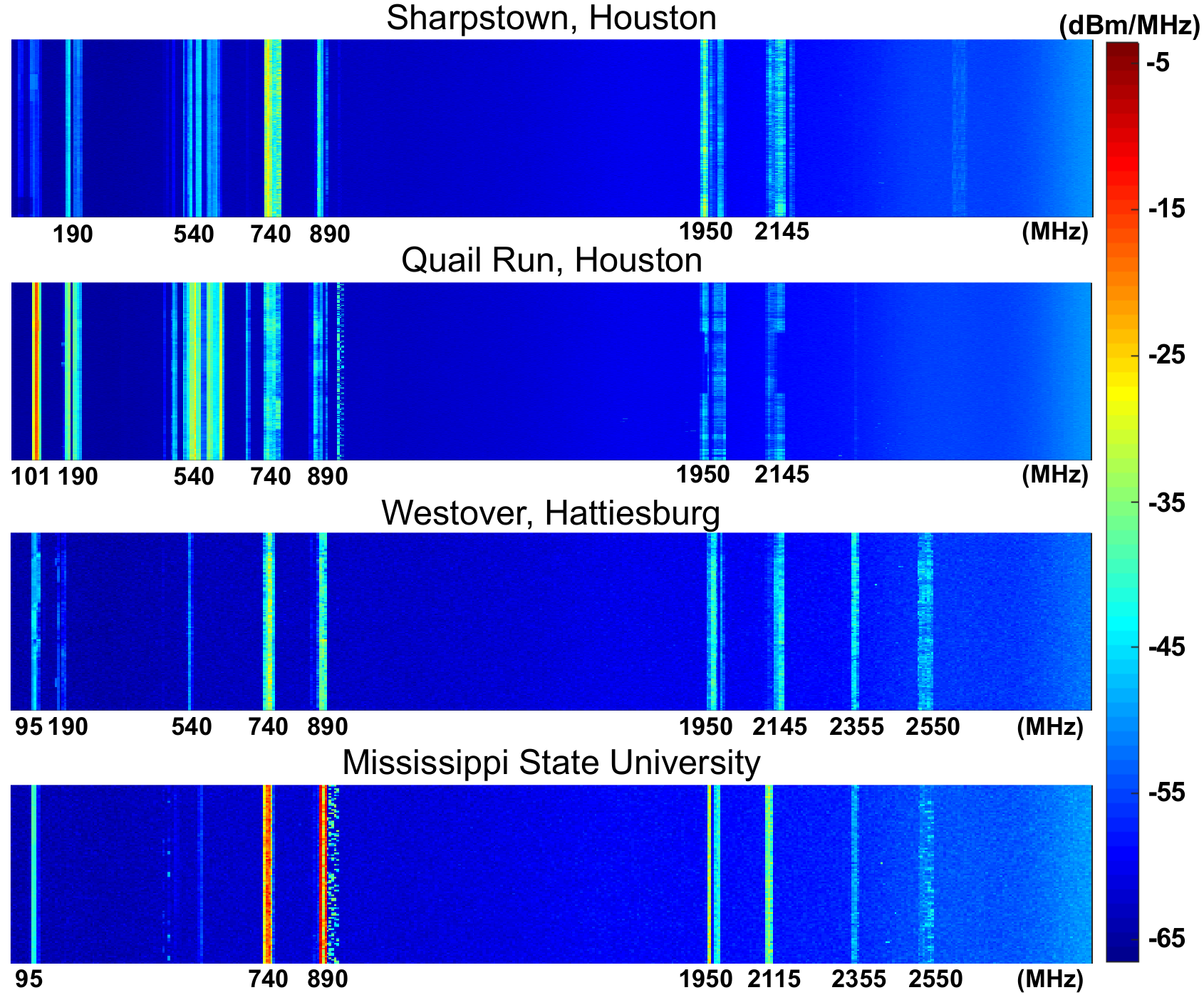}}
  \caption{Spectrograms of ambient RF energy in outdoor environments.}\label{fig:Spectrum}
\end{figure}

In Fig.~\!\ref{fig:Spectrum}, we show the spectrograms of renewable RF energy measured at different locations using the Keysight N9340B spectrum analyzer and the Keysight N6850A broadband omnidirectional antenna. Each measurement took around 15 minutes. The X-axis represents the frequency and the Y-axis is time, with the trace from the latest sweep displayed at the bottom of the figure; additionally, the amplitude in each $1$\,MHz frequency band is represented by color. As illustrated in the figure, the power density of radio waves is typically low (less than $-35$\,dBm/MHz) at most frequencies. Therefore, not all renewable RF energy can be successfully received through sustainable IoT devices. In a certain frequency band, only the intensity of the incident power is higher than the sensitivity of an RF harvester, usually above $-25$\,dBm (the yellow and orange lines in the figure), can activate the harvester~\cite{valenta2014harvesting}.

By comparing the first two spectrograms in Fig.~\!\ref{fig:Spectrum}, it can be observed that due to the high propagation attenuation of radio waves and the non-uniform deployment of RF sources (e.g., TV tower, radio base station, and WiFi access point), the power density of renewable RF energy at different locations in a city is completely different. For instance, the average power density at FM frequency band (around $101$\,MHz) can achieve $-16.7$\,dBm in Quail Run, but it is only $-56.3$\,dBm in Sharpstown. Furthermore, the RF intensity measured at the same location but at different frequency bands may have significant differences. For instance, in Mississippi State University, the strength of RF signal at 3G frequency band ($890$\,MHz) can be above $-10$\,dB, which is $20$\,dB higher than that measured at 4G LTE band ($1950$\,MHz).

\begin{figure}[htb]
\centerline{\includegraphics[width=8.7cm]{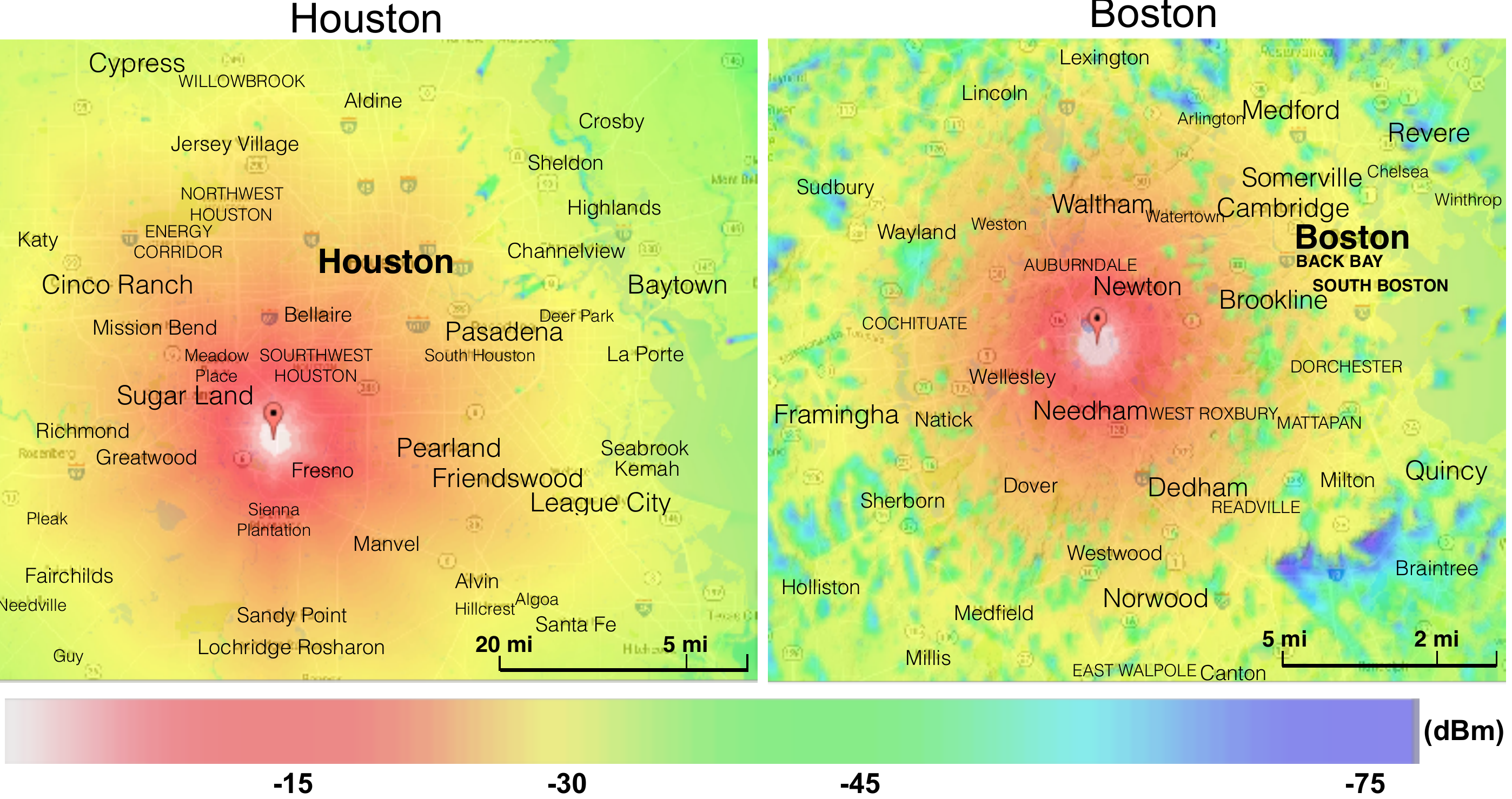}}
  \caption{The effect of terrain on RF power density distribution. Note that the map on the right (Boston) is four times the map on the left (Houston). The effective radiated power (EPR) and the height above average terrain (HAAT) of the TV station in Houston are $1$\,kW and $580$\,m, respectively; the television station in Boston has two similar parameters, respectively $1.35$\,kW and $390$\,m.}
  \label{fig:SpectrumDis}
\end{figure}

Even if RF energy harvesters have the same distance to an energy source, their energy harvesting rates can be different. This is because the propagation attenuation of the radio wave is not uniform, but is affected by weather, obstacles, and terrain. In Fig.~\!\ref{fig:SpectrumDis}, we use the tool provided in \cite{tv2019tv} to plot the spatial distribution of TV signals in two different cities. As illustrated in the figure, the signal attenuation in the plain area like Houston is relatively slow and smooth; therefore, an RF harvester 5 miles from the TV tower can still get $-15$\,dBm of incident power. In contrast, the spatial distribution of radio energy fluctuates drastically in the hilly area like Boston. For an energy harvester 5 miles from the TV tower, the intensity of the incident power can be as low as $-40$\,dBm.

Based on the above analysis, it can be appreciated that although we are surrounded by renewable RF energy, their power density is generally weak and their distribution in frequency and space may be uneven. Therefore, it is important to investigate the radio environment in the target area before deploying sustainable IoT devices. In the next section, we study how to model the radio energy harvesting process in outdoor environments.

%% file: EngHarModel.tex
\section{Radio Energy Harvesting Process}
\label{sec:EngHarPro}
In this section, we briefly introduce three factors that directly affect the efficiency of RF energy harvesting. After that, we use supercapacitors as an example to show the charging characteristics of real radio energy harvesters.
  
\subsection{Factors that Affect Harvesting Efficiency}
\label{subsec:Character}
An RF energy harvester consists of three key components: an impedance matching circuit, a rectifier, and an ESC. The matching circuit adjusts the output impedance of the receiving circuit to be the same as the input impedance of the antenna, thereby maximizing the energy conversion efficiency. The rectifier converts alternating current (AC) to direct current (DC) and then boosts the output voltage appropriately to charge the ESC. The collected energy is then stored in the ESC to power the target device. 

Radio energy powered IoT devices charge and discharge their ESCs more frequently than battery powered wireless nodes. In addition, IoT devices need to work reliably in harsh outdoor environments with high or low temperatures. Therefore, when selecting the ESC, a supercapacitor having a long lifespan ($100,000$ to a million charging cycles) and a wide range of operating temperatures ($-40^{\circ}$F to $248^{\circ}$F) is more popular than a rechargeable battery~\cite{hung2009wide}. To this end, this paper mainly uses supercapacitors as an example to analyze the energy harvesting process of RF energy powered IoT devices.

The following three factors can significantly affect the efficiency of radio energy harvesting:
\vspace{0.1cm}
\begin{adjustwidth}{-0.77cm}{0cm}
\begin{description}
\setlength{\labelsep}{-0.95em}
\itemsep 0.07cm
  \item[\emph{a)}] \emph{Frequency of radio waves:} If the frequency of the radio energy does not match the resonant frequency of the antenna, the harvester cannot receive energy efficiently because the incident radio waves will be reflected back to the air. Radio energy harvester built with Schottky diodes can achieve the energy conversion efficiency of $50\%$ or more when the frequency of the input energy matches the circuit~\cite{sun2012design, powercast2016p2110b, agrawal2014realization}. 
  \item[\emph{b)}] \emph{Intensity of incident power:} In order for a harvester to receive energy efficiently, the intensity of the RF power should be within a reasonable range. If the incident energy is too weak, the voltage swing at diodes of the energy harvesting circuit is lower than or equal to the forward voltage. In this case, the diode is turned off and only a small current can flow through the diode, resulting in low energy harvesting efficiency. If the incident energy is too strong, the diode is reverse-conducting and the current under reverse bias becomes larger as the voltage swing of the diode exceeds the breakdown voltage. In this case, the energy collection efficiency is also low.
  \item[\emph{c)}] \emph{ESC's energy level:} Even if the outdoor RF environment is stable for a long time, the energy harvesting rate is not a constant due to the nonlinear charging characteristic of the ESC. As analyzed in \cite{luo2018revisiting}, the charging speed is fast when the ESC is empty, and then decreases as the energy level becomes high. Therefore, the  energy harvesting rate is determined not only by the RF power density, but also by the current energy level of the ESC.
\end{description}
\end{adjustwidth}
\vspace{0.1cm}

It is worth noting that the open-circuit voltage of the RF energy harvester is not a constant, but varies greatly with the strength of the incident radio power. This output characteristic is very different from that of a solar harvester, which can maintain fairly stable open-circuit voltage around $0.58$\,V from dull to bright sunlight. Denote the open-circuit voltage and the intensity of the incident power by $V_{OC}$ and $P_i$, respectively. In Table~\ref{tab:OCV}, we use the Powercast P2110 energy harvester as an example to show how $V_{OC}$ rises as $P_i$ increases. Similar results can be observed in most existing RF energy harvesters~\cite{nintanavongsa2012design, papotto201190, jabbar2010rf}. 

\begin{table}[htp]
\footnotesize
\centering
\caption{$V_{OC}$ of P2110 energy harvester with respect to $P_i$.}
\label{tab:OCV}
\begin{tabular}{|
>{\columncolor[HTML]{ECF4FF}}c |
>{\columncolor[HTML]{C0C0C0}}c |
>{\columncolor[HTML]{EFEFEF}}c |
>{\columncolor[HTML]{C0C0C0}}c |
>{\columncolor[HTML]{EFEFEF}}c |
>{\columncolor[HTML]{C0C0C0}}c |
>{\columncolor[HTML]{EFEFEF}}c |
>{\columncolor[HTML]{C0C0C0}}c |}
\hline
$P_i$ (dBm)                          & -14 & -11.3 & -8.5 & -7 & -5  & -3  & -2 \\ \hline
\cellcolor[HTML]{DAE8FC}$V_{OC}$ (V) & 0.4 & 0.9   & 1.6  & 2  & 2.6 & 3.2 & 4  \\ \hline
\end{tabular}
\end{table}

\subsection{Charging Characteristics of RF Energy Harvester}
\label{subsec:Char}
\begin{figure}[htb]
\centerline{\includegraphics[width=6cm]{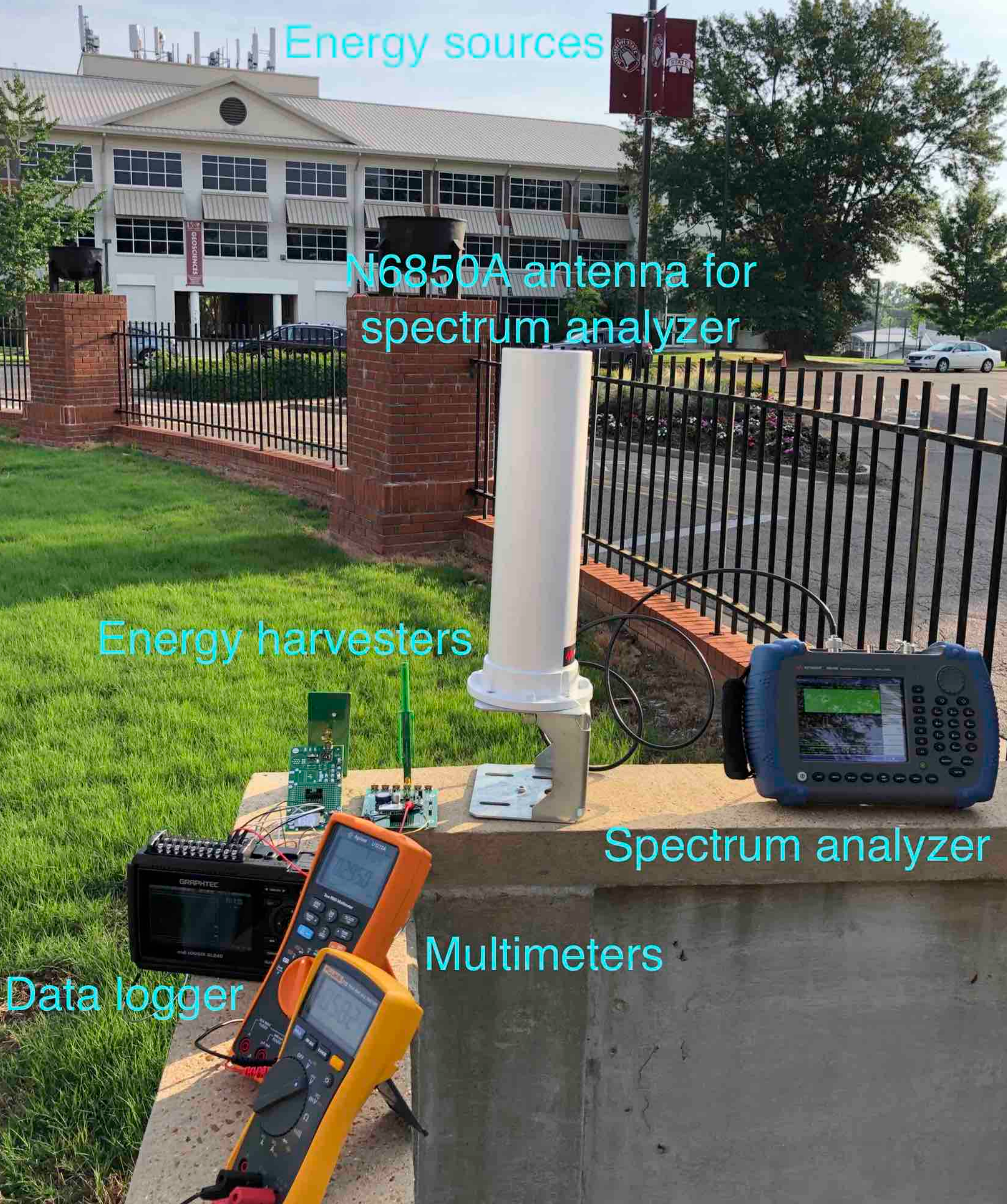}}
  \caption {Outdoor experiment conducted at Mississippi State University. The distance between energy source and harvester is about 280 feet.}
  \label{fig:Scenario}
\end{figure}

Due to the low power density of renewable RF energy in the air, the intensity of the incident power at an IoT device is typically less than $0$\,dBm. However, commercial transceivers\footnote{Here, we are talking about active transceivers. They do not include backscattering transmitters and near-field inductively coupled transmitters, which are passive systems that are commonly used in RFID and NFC devices for short-range communications.}  consumes more than a few milliwatts of power during data transmission and reception~\cite{microchip2016atmel, cc2013cc2500}. As a result, the duty cycle of IoT devices is usually low to accumulate enough energy for short-term communication. To study the characteristics of renewable RF energy harvesting process, we conducted a large number of outdoor experiments. A typical experiment scenario can be found in Fig.~\!\ref{fig:Scenario}. 

\begin{figure}[htb]
\centerline{\includegraphics[width=8.0cm]{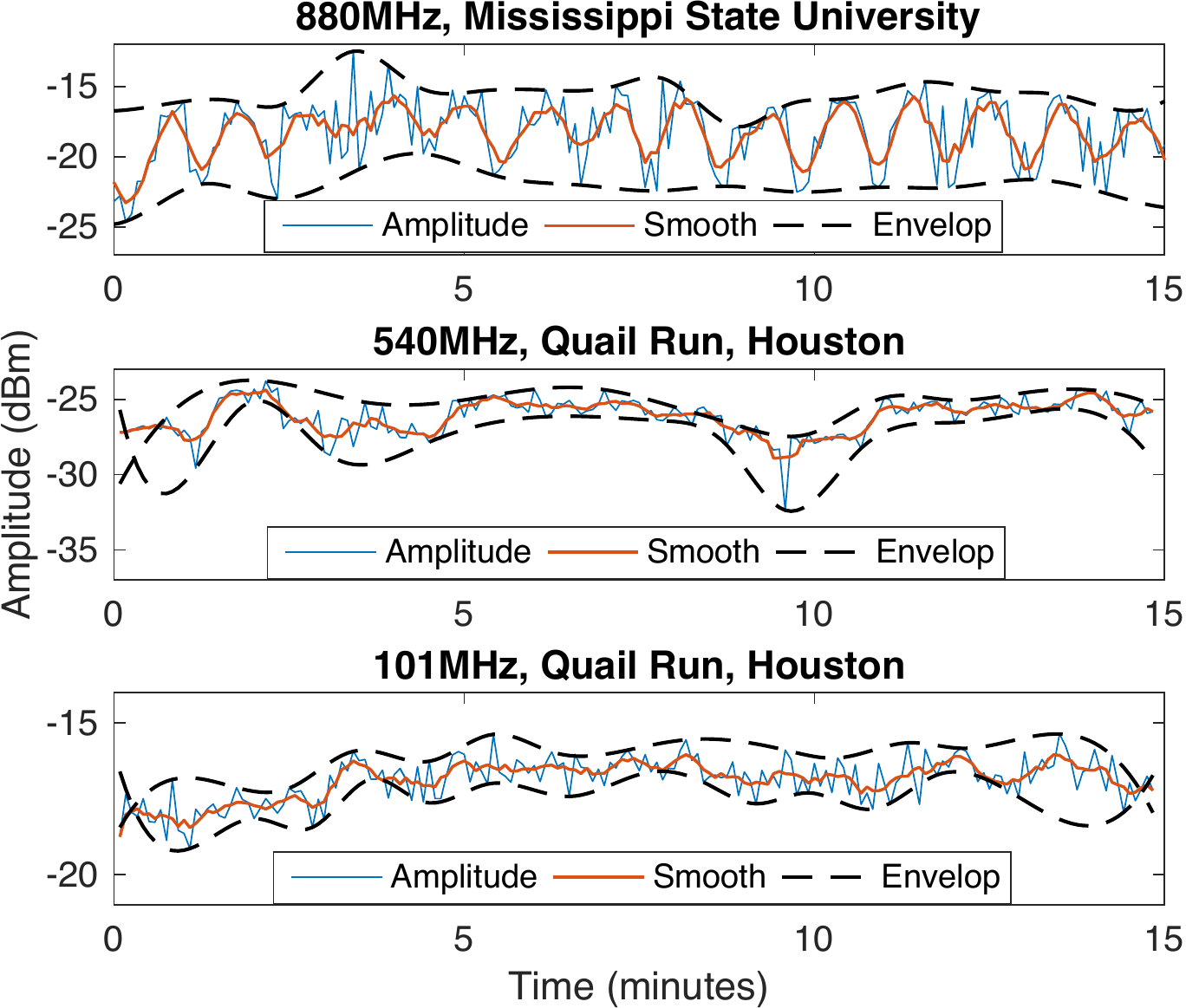}}
  \caption {Fluctuations of RF energy over time.}
  \label{fig:RF_amplitude}
\end{figure}

How the strength of ambient RF energy changes in outdoor environments is illustrated in Fig.\!~\ref{fig:RF_amplitude}, where the time interval between neighboring measurements is $5.75$\,s. In the figure, the envelop of radio waves are also printed to visualize the upper and lower bounds in the energy strength.  As can be observed from the figure, the instantaneous RF intensity has random and large fluctuations (over $5$ to $10$\,dBm) in each frequency band. As a result, the variance of RF amplitude is high, and the smoothed power intensity and the envelop of the renewable radio energy change slowly with time. Similar results are also reported in \cite{muncuk2018multiband} and \cite{pinuela2013ambient}, where RF power density is measured in Boston and London, respectively. This indicates that random fluctuations are an inherent feature of renewable radio energy in outdoor environments.

Intuitively, the energy harvesting rate of sustainable IoT devices should fluctuate randomly with RF power density because the open-circuit voltage depends on the energy intensity as measured in Table~\ref{tab:OCV}. That is why existing works commonly assume that the renewable radio energy received by the harvester is discrete energy packets with random sizes and arrival time. However, our next experiment results will challenge this assumption.

To model the energy harvesting process in the real world, we deployed a Powercast P2110 harvester at Mississippi State University to receive renewable RF energy radiated from a  $880$\,MHz cellular base station. The density of the RF energy at the harvester can be found in Fig.\!~\ref{fig:Spectrum} and Fig.\!~\ref{fig:RF_amplitude}. In the experiment, the P2110 received energy through a $1$\,dBi omnidirectional antenna and then stored the energy in two different supercapacitors.

\begin{figure}[htb]
\centerline{\includegraphics[width=8.7cm]{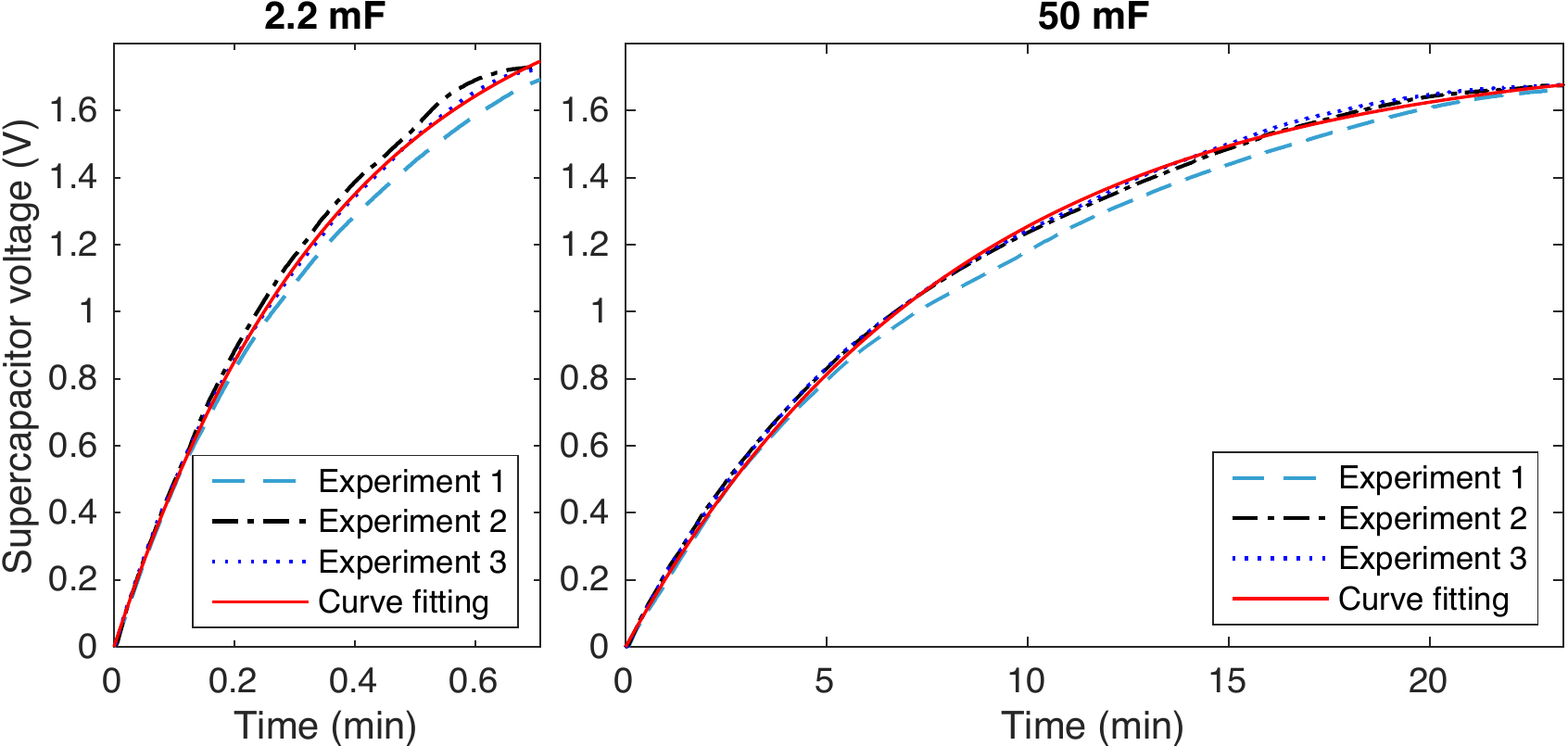}}
  \caption {Charging curve of a renewable energy harvester in the real world. The capacitance of supercapacitors: $2.2$\,mF (left), and $50$\,mF (right).}
  \label{fig:CharProc}
\end{figure}

In Fig.~\!\ref{fig:CharProc}, we show the ESC charging curve for multiple measurements. As can be observed from the figure, the result is similar to the charging process of a capacitor with a stable charging voltage. In addition, in each of the figures, all of the charging curves almost overlap each other, indicating a good consistency of the charging process measured at different times. These observations are counter-intuitive because they show that \emph{even if the intensity of the incident energy fluctuates significantly in outdoor environments, the ESC voltage of the RF energy harvester will rise steadily over time}.

A smooth and consistent charging curve provides an opportunity for IoT devices to accurately predict their future energy harvesting process from a small number of historical measurements. This can be verified by simplifying the RF harvester to a resistor\,--\,capacitor (RC) circuit, and then using the capacitor charging equation to describe the energy harvesting process:  
\begin{equation}
\label{eq:4-01}
	V_{cc} (t) = V_{OC}\left(1-e^{\frac{-t}{RC}}\right),
 \end{equation}
where $V_{cc}(t)$ is the voltage level of the ESC at time $t$, $V_{OC}$ is the open-circuit voltage of the energy harvester, $R$ is the equivalent impedance of the harvesting circuit, and $C$ is the capacitance of the supercapacitor.

To predict the entire charging curve, an IoT device can measure its ESC voltage multiple times at the beginning of the charging process and then apply the least-squares fitting\footnote{In some energy harvesters, the ESC can be temporarily disconnected from the main circuit by an electronic switch, and the IoT device can directly measure the open-circuit voltage. In this case, the device only needs to calculate $R$, which significantly reduces the computational complexity of the least-squares fitting.} to calculate the unknown variables $R$ and $V_{OC}$ in (\ref{eq:4-01}). Afterward, the amount of energy stored in the supercapacitor at time $t$, which is denoted by $E_s(t)$, can be obtained via $E_s(t)\!=\!\frac{1}{2}CV^2_{cc}$.

In Fig.~\!\ref{fig:CharProc}, we compare the results of least-squares fitting in (\ref{eq:4-01}) with the experiment  data. From the figure, it can be observed that the ESC voltage predicted by the least-squares fitting matches the true value very well. To further evaluate the performance of the charging prediction, we define the prediction error as the difference between the predicted ESC voltage and the average of the experiment measurements.  The results show that the mean values of the prediction errors are very small, which are only $5.1\!\times\!10^{-3}$\,V and $8.4\!\times\!10^{-3}$\,V when $C\!=\!2.2$\,mF and $C\!=\!50$\,mF, respectively. 

It is worth nothing that the equivalent impedance of the harvesting circuit is not only affected by the intensity of the incident energy but also by the capacitance of the supercapacitor. Using Fig.~\!\ref{fig:CharProc} as an example, when capacitances of the supercapacitors are $2.2$\,mF and $50$\,mF, the values of $R$ calculated by the least-squares fitting are $170.6$\,$\Omega$ and $3.7$\,k$\Omega$, respectively; the latter is $21.7$ times higher than the former. In fact, it is difficult to accurately simulate the charging process in a dynamic RF environment based on the hardware design of the energy harvester and the input of RF energy. However, if we choose the least-squares fitting, the IoT device can treat the harvesting circuit and environment as a black box, and the ESC voltage is the only information needed to predict the charging curve.

%% file: EngSingle.tex
\section{Energy Consumption of Single Packet}
\label{sec:EngSig}
This section will focus primarily on the energy consumption of single-packet transmission. We first analyze the overhead energy generated by the communication protocol and the hardware of an IoT device. Then, we study the relationship amongst system power consumption, transmission power, and supply voltage.

\subsection{Overhead Energy of Single-Packet Transmission}
\label{subsec:OverEng}
During data transmission, IoT devices inevitably generate some overhead energy, the amount  of which depends mainly on the hardware design and the protocol stack running in the system. Next, we use ATmega256RFR2 microcontroller-based IoT device running the IEEE 802.15.4 wireless communication protocol as an example to model the overhead energy of single packet transmission. Our analysis method is universal, so it can be easily extended to other IoT devices and communication protocols.

\begin{figure*}[htp]
\centering
\includegraphics[width=18cm]{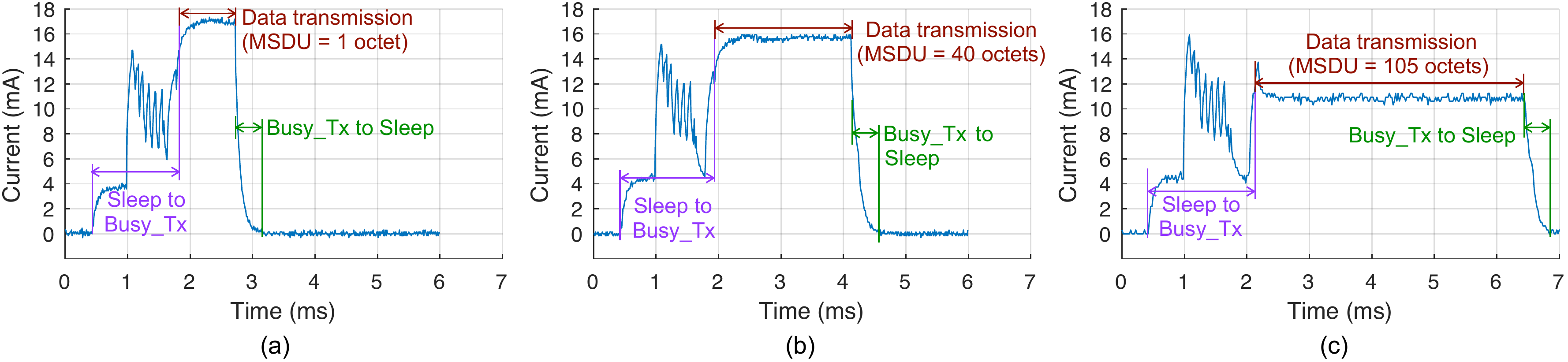}
\caption{Current consumptions of ATmega256RFR2 under different configurations. (a) $P_t=3.5$\,dBm, $L_e=1$ octet, and $V_{cc}=2$\,V; (b) $P_t=1.8$\,dBm, $L_e=40$ octets, and $V_{cc}=2.5$\,V; (c) $P_t=-16.5$\,dBm, $L_e=105$ octets, and $V_{cc}=3.5$\,V.}
\label{fig:PowerWav}
\end{figure*}

Based on our measurements, ATmega256RFR2 consumes as little as $1.08\,\mu$W ($0.6\,\mu$A$\times 1.8$\,V) of power in the deep sleep mode, which is affordable for renewable radio energy powered IoT devices. However, the power consumption of the activated microcontroller will increase to at least $7.2$\,mW ($4$\,mA$\times 1.8$\,V), which in most cases is much higher than the energy harvesting rate, as described in Section~\ref{sec:EngHarPro}. As a result, the sustainable IoT device may not be able to work continuously in an outdoor environment. It needs to be dormant most of the time and will only be woken up by an interrupt when a specific event is detected.

Each time the IoT device is woken up, the system needs some energy to ramp up the voltage regulator and initialize the phase-locked loop (PLL). In addition to the hardware, the communication protocol also consumes a certain amount of energy to transmit some functional data for packet control, synchronization, frame checking, and encryption. In this paper, we consider the MAC layer payload, i.e., the MAC service data unit (MSDU), as the effective data; all other functional frames, including the synchronization header (SHR), physical header (PHR), MAC header (MHR), and frame check sequence (FCS), are overhead.

To measure the overhead energy introduced by the hardware and communication protocol, we set the MAC addressing fields (AF) and the auxiliary securing header (ASH) in IEEE 802.15.4 to $6$ octets and $10$ octets, respectively. In this case, the MSDU payload will be between $0$ and $106$ octets. When running IEEE 802.15.4 at the PHY and MAC layers, the first six octets ($5$ octets of SHR and $1$ octet of PHR) must be transmitted at a rate of $250$\,kbps; the remaining octets can then be sent at a customized data rate. We use $L_{OP}$ to represent the length of the overhead frames in the physical layer service data unit (PSDU). Then $L_{OP}\!=\!21$ octets, including the frame control field (FCF, $2$\,octets), sequence number (SN, $1$\,octet), AF ($6$\,octets), ASH ($10$\,octet), and FCS ($2$\,octets).

Denote the transmission power, supply current, and data rate by $P_t$, $C_c$, and $r_d$, respectively. Let $V_{cc}$ be the power supply voltage. The size of the effective data, i.e., the MSDU payload, is represented by $L_e$. The wake-up time represented by $T_w$ is defined as the time it takes for the IoT device to switch from deep sleep to data transfer mode. Let the sleep time represented by $T_s$ be the time it takes for the device to return from the data transfer mode to the deep sleep mode. 

In Fig.~\!\ref{fig:PowerWav}, we set $r_d\!=\!250$\,kbps and then measure the supply current of ATmega256RFR2 under three different combinations of $L_e$, $P_t$, and $V_{cc}$. The key features of the energy overhead observed from the figure are summarized as follows: 
\vspace{0.1cm}
\begin{adjustwidth}{-0.77cm}{0cm}
\begin{description}
\setlength{\labelsep}{-0.95em}
\itemsep 0.07cm
  \item[a)]  Comparing the wake-up time under different settings, it can be realized that $T_w$ increases when the MSDU payload becomes large. As illustrated in Fig.~\!\ref{fig:WakeupTim}, the wake-up time is a linear function of the MSDU payload; therefore, $T_w$ can be accurately estimated as follows:
 \begin{equation}
\label{eq:5A_01}
  T_w = 0.004\,L_e+1.395,
\end{equation}
where $L_e$ is the MSDU payload in terms of octets. From (\ref{eq:5A_01}), we know that reducing $L_e$ can wake up the system faster while reducing the energy consumption. For instance, when the MSDU payload is full ($106$ octets), ATmega256RFR2 takes $1.82$\,ms to wake up; when the payload is only $1$ octet, this time can be reduced to $1.4$\,ms, $23\%$ shorter than the former.

\begin{figure}[htb]
\centerline{\includegraphics[width=6.5cm]{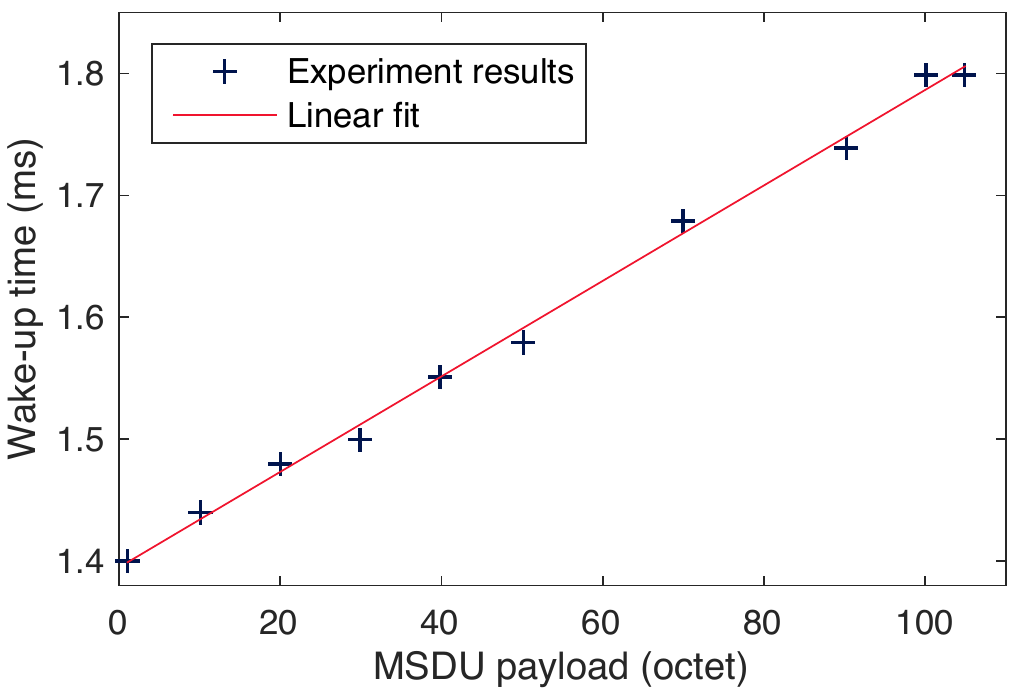}}
\caption {The wake-up time as a function of the MSDU payload.}
\label{fig:WakeupTim}
\end{figure}

   \item[b)]  The average current consumption of the microcontroller from deep sleep to data transmission is $7.8$\,mA. It does not change with $V_{cc}$ and $T_w$. As a result, the wake-up energy represented by $E_w$ becomes a linear function of $V_{cc}$. According to (\ref{eq:5A_01}), we have that 
\begin{equation}
\label{eq:5A_02}
\begin{array}{lll}
\vspace{0.17cm}	
	E_w \!=\!7.8V_{cc}T_w=\! 7.8V_{cc}\,(0.004 L_e+1.395),
\end{array}
\end{equation}
where the units of $E_w$ and $V_{cc}$ are $\mu$J and $V$, respectively.

   \item[c)] As shown in Fig.~\!\ref{fig:PowerWav}, the sleep time of the IoT device, $T_s$, is a constant. It is always $0.45$\,ms and does not change with the supply voltage and MSDU payload. In addition, the average current consumption of the device switching from the data transfer mode to the deep sleep state is approximately half of the current during data transmission. If we use $E_{ds}$ to represent the system energy consumption during $T_s$, then it can be obtained that
 \begin{equation}
\label{eq:5A_03}
	E_{ds}= \displaystyle\frac{V_{cc}T_sC_c}{2}= 0.225\,V_{cc}\,C_c,
\end{equation}
where the units of $E_{ds}$, $V_{cc}$, and $C_c$ are $\mu$J, V, and mA, respectively.
  \item[d)] As illustrated in Fig.~\!\ref{fig:PowerWav}(a), even if the MSDU payload is only $1$ octet, the transceiver will take some time to send the entire packet due to the overhead frames added by the communication protocol. Let  $L_{SHR}$ and $L_{PHR}$ be the lengths of SHR and PHR, respectively. Then for $L_e\!\in\![0,106]$, the total transmission time of the data packet represented by $T_d$ can be calculated as follows:
\begin{equation}
\label{eq:5A_04}
\begin{array}{lll}
\vspace{0.17cm}	
	T_d \!\!\!\!&=&\!\!\! \displaystyle\frac{L_{SHR}+L_{PHR}}{250\,\text{kbps}}+\displaystyle\frac{L_{OP}+{L_e}}{r_d}\\
	\!\!\!&=&\!\!\! 192\,\mu s+\displaystyle\frac{\text{168\,bit}+{L_e}}{r_d}.
\end{array}
\end{equation}
It can be obtained from (\ref{eq:5A_04}) that when the MSDU payloads are $1$ octet and $106$ octets, the effective data transmission time is only accounts for $3.6\%$ and $80\%$ of the total transmission time.
\end{description}
\end{adjustwidth}
\vspace{0.1cm}

If the data to be sent is greater than the maximum MSDU payload, the data will be divided into several consecutive packages. The IoT device does not enter deep sleep mode between the transmission intervals of two consecutive packets. It just turns off the transceiver but keeps the CPU active to avoid repeated wake-ups. Nevertheless, when switching between the transceiver off and data transfer modes, the IoT device still generates some overhead energy, which will be studied in Section~\ref{sec:EngCon}.

\subsection{System Power Modeling}
\label{subsec:SysPow}
In order to efficiently manage the received energy, we need to know the relationship between the system power consumption and the transmission power. For an IoT device, the average power of the processor and peripherals is stable during data transmission~\cite{microchip2016atmel}; therefore, the system power consumption is usually assumed to be the transmission power of the transceiver plus a constant system power. However, the power consumption in real systems is not that simple. Next, we propose a general power consumption model for different IoT devices.

\begin{figure}[htb]
\centerline{\includegraphics[width=7cm]{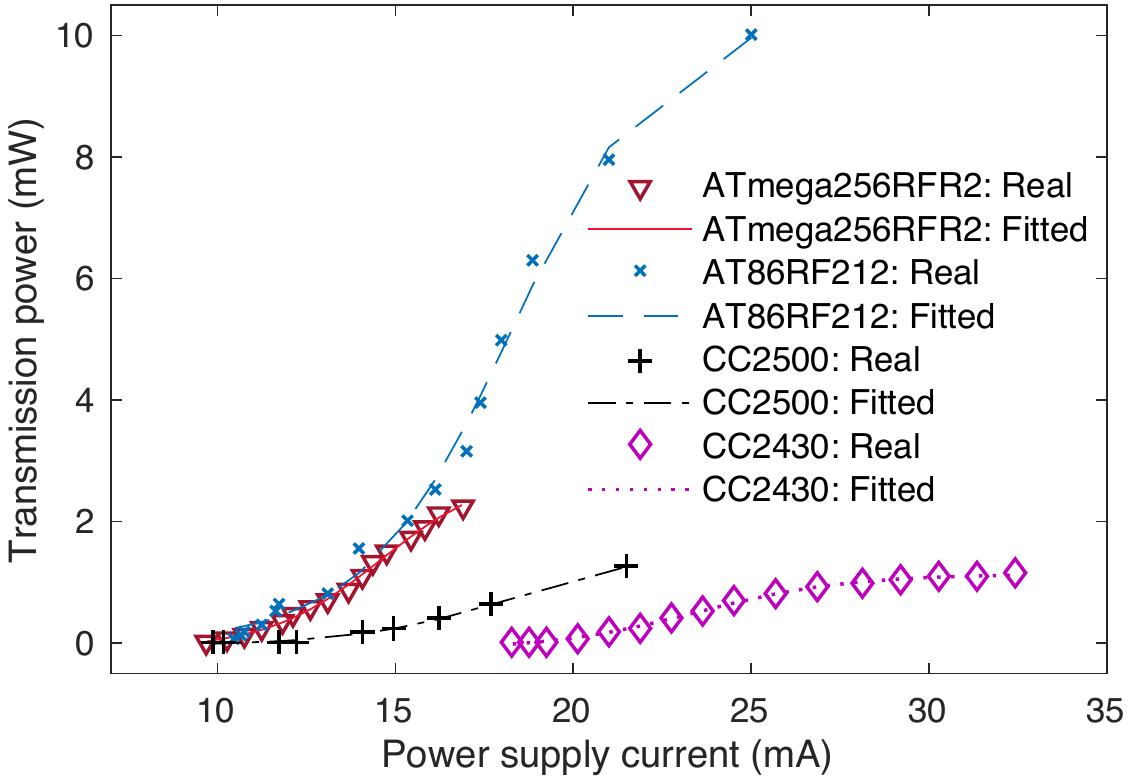}}
  \caption {The transmission power versus the system total power consumption when the supply voltage is $1.8$\,V.}
  \label{fig:PowerFit}
\end{figure}

In Fig.~\!\ref{fig:PowerFit}, we fix the power supply voltage of each IoT device to $1.8$\,V, and then show the \textcolor{blue}{transmission power} ($P_t$) as a function of the supply current ($C_c$). As shown, $C_c$ does not increase linearly with $P_t$. Specifically, when an IoT device is activated but there is no data to send, i.e., when $P_t\!=\!0$, only the processor and peripherals consume little current, which is less than $10$\,mA in most cases. As $C_c$ becomes large, $P_t$ increases rapidly and forms an S-shaped curve in the figure. Based on this observation, we can construct a modified sigmoid function to model $P_t$ as follows:
\begin{equation}
\label{eq:5B-01}
	P_t = \alpha_1-\displaystyle\frac{\alpha_2}{e^{\alpha_3(C_c-\alpha_4)}+1},
 \end{equation}
where $\alpha_1$ to $\alpha_4$ are scaling and displacement coefficients. To calculate the scaling and displacement coefficients, the IoT device needs to change its transmission power by assigning different values to the transmission register while monitoring the system power consumption (the product of the supply voltage and the supply current) at different transmit powers. The least-squares fitting can then be applied to calculate the scaling and displacement coefficients in (\ref{eq:5B-01}).

In Fig.~\!\ref{fig:PowerFit}, we compare the profiles of $P_t$ obtained from (\ref{eq:5B-01}) with $P_t$ from real systems. As observed from the figure, the results obtained from (\ref{eq:5B-01}) and the data collected from different IoT devices match very well. This verifies that the sigmoid function constructed in (\ref{eq:5B-01}) can accurately describe the relationship between the system power consumption and the transmission power of the IoT device.

Eventually, substituting $C_c \!=\! P_s/V_{cc}$ into (\ref{eq:5B-01}), we have that 
\begin{equation}
\label{eq:5B-02}
	P_t = \alpha_1-\displaystyle\frac{\alpha_2}{e^{\alpha_3\left(\frac{P_s}{V_{cc}}-\alpha_4\right)}+1}.
 \end{equation}
In (\ref{eq:5B-02}), we described the relationship among the transmission power ($P_t$), the supply voltage ($V_{cc}$), and the system power consumption ($P_s$). With (\ref{eq:5B-02}), we confirm that the system power consumption is a modified sigmoid function of the transmission power.

%% file: EngContinuous.tex
\section{Energy Consumption of Continuous Packets}
\label{sec:EngCon}
In this section, we will focus on the energy consumption of IoT devices during continuous data transmission. We first introduce how to calculate energy consumption after considering the reduction of the supply voltage during data transmission. After that, the overhead energy of sending adjacent packets is analyzed.

\subsection{Energy Consumption for Continuous Packets Transmission}
\label{subsec:EngConTran}
Due to the thin power density of renewable RF energy and the size as well as weight constraints of IoT devices, the ESC capacity is usually low. As a result, the ESC voltage, which is the also the supply voltage of the IoT device, may vary considerably with energy consumption and energy reception in a short time. For this reason, we concern about whether the system power consumption will change significantly with the ESC voltage given a transmission power.

\begin{figure}[htb]
\centerline{\includegraphics[width=7cm]{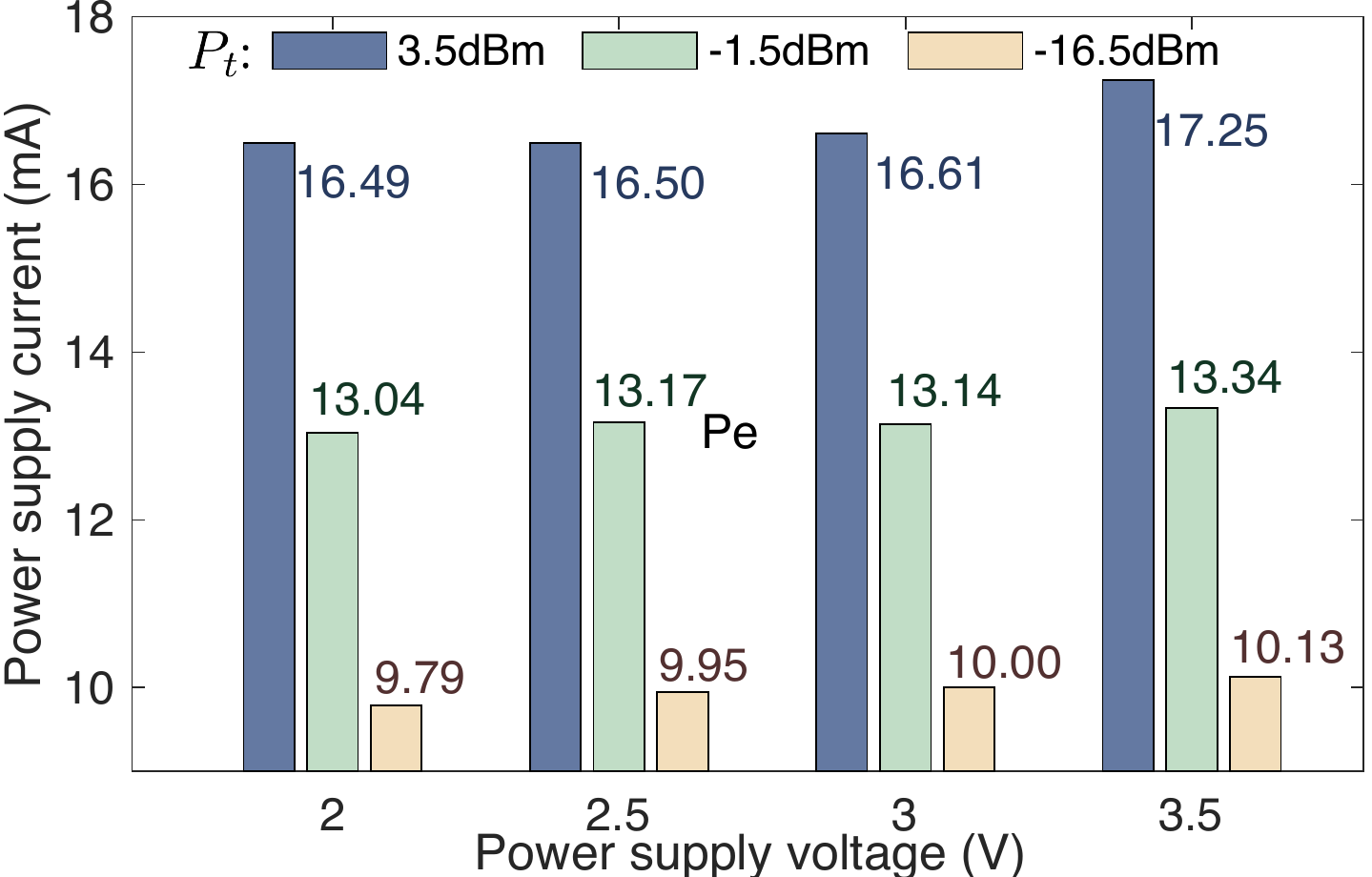}}
  \caption {Variation of $C_c$ with respect to $V_{cc}$.}
  \label{fig:Curr_Vol}
\end{figure}

To answer the above question, we let ATmega256RFR2 microcontroller send data with constant power, and then show how $C_c$ changes with $V_{cc}$ in Fig.~\!\ref{fig:Curr_Vol}. As can be observed from the figure, given a transmission power ($P_t$), the supply current just increases slightly when the supply voltage becomes high. Specifically, when $V_{cc}$ is increased from $2$\,V to $3.5$\,V, $C_c$ increases by only $4.6\%$, $5.4\%$, and $3.5\%$ at $P_t=3.5$\,dBm, $-1.5$\,dBm, and $-16.5$\,dBm, respectively. This indicates that the effect of the varying supply voltage on the supply current is negligible. Therefore, the system power consumption, $P_s$, increases linearly with $V_{cc}$ because it is the product of $C_c$ and $V_{cc}$. In other words, reducing the supply voltage can decrease the power consumption of data transmission.

From the above analysis, we realize that given a transmission power and a data rate, the energy consumption of continuous data transmission is not a fixed value, but gradually decreases with time. This is because the ESC voltage of the IoT device drops rapidly during data transmission. This issue has not received much attention in existing energy models and power management strategies. Next, we will analyze how system power consumption decreases with data transmission, and the results will help us build an accurate energy model for RF energy powered IoT devices.

Through a simply derivation given in the Appendix~\!\ref{subapp:EngCon}, we know that the energy consumed when transmitting data bit $i$ of packet $j$, which is represented by $E_s^{i, j}$, can be calculated as follows:
\begin{equation}
\label{eq:6A-01}
  E_s^{i, j} = E_s^{1, j}- \displaystyle\frac{(i-1)}{C}\!\!\left(\frac{C^j_c}{r^j_d}\right)^2,
\end{equation}
where $r^j_d$ is the data rate of packet $j$; $V_{cc}^{1, j}$ is ESC voltage before sending the first bit of packet $j$; $C^j_c$ is the supply current when transmitting packet $j$; $E_s^{1, j}=V_{cc}^{1, j}\,C^j_c\,/r^j_d$ is the energy consumed to send the first bit of packet $j$. We call $V_{cc}^{1, j}$ as the initial voltage.

It can be observed from (\ref{eq:6A-01}) that if the ESC capacity is small, $E^{i, j}_s$ rapidly decreases as the number of bits increases. Specifically, assume the MSDU payload of IEEE 802.15.4 protocol is full, then the size of effective data is $127$ octets, i.e., $1016$ bits. Assume the capacity of the supercapacitor is $C=0.12$\,mF; the power supply current of ATmega256RFR2 is $C_c=16.24$\,mA to send data at $3.5$\,dBm; the data rate is $r_d=250$\,kbps; the initial voltage of the supercapacitor is $V^1_{cc}=2.5$\,V. According to (\ref{eq:6A-01}), it can be obtained that the energy consumption of the first bit and the last bit of a data packet are $0.162\,\mu$J, and $0.127\,\mu$J, respectively; the latter is $21.6\%$ lower than the former. This indicates that if the IoT device uses a low-capacity ESC, the energy consumption is significantly reduced during data transmission. This effect cannot be ignored in the power management strategy.

\subsection{Overhead Energy Between Continuous Packets}
\label{subsec:OverEngCon}
\begin{figure}[htb]
\centerline{\includegraphics[width=7cm]{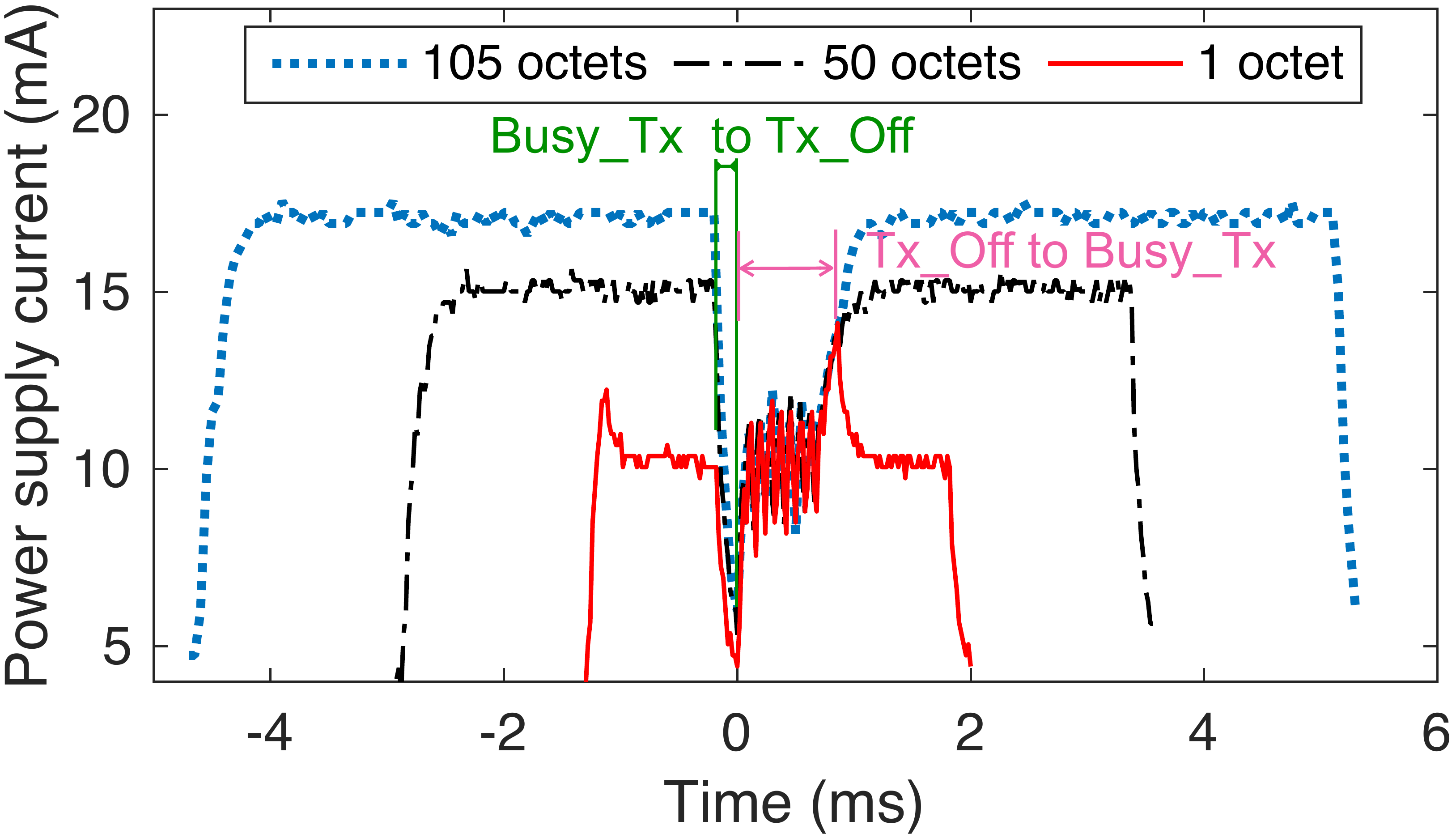}}
  \caption{The waveforms of power supply current when two consecutive packets are sent. Red curve: $L_e=1$\,octet, $V_{cc} \equiv 2$\,V, $P_t=-16.5$\,dBm; black curve: $L_e=50$\,octets, $V_{cc} \equiv 3.5$\,V, $P_t=1.2$\,dBm; $L_e=105$\,octets, $V_{cc} \equiv 2.5=2.5$\,V, $P_t=3.5$\,dBm.}
  \label{fig:Overhead}
\end{figure}

If an IoT device is scheduled to send multiple packets in one active cycle, it will not enter the deep sleep mode until the final packet is sent. To investigate the overhead energy between transmissions of adjacent packets, we let the ATmega256RFR2 microcontroller send two consecutive packets at different settings and then show the supply current waveform in Fig.~\!\ref{fig:Overhead}.

As illustrated in Fig.~\!\ref{fig:Overhead}, the microcontroller can turn off the transceiver in $0.2$\,ms; however, it takes a relatively long to turn on the transceiver. When the transceiver is turned off, only the CPU consumes energy. In this case, the supply current drops to 4 mA. According to the measurement results, the microcontroller needs to spend $0.86$\,ms of time and $10.25$\,mA of supply current to turn on the transceiver; these two parameters are constants independent of the transmission power ($P_t$), MSDU payload ($L_e$), and supply voltage ($V_{cc}$). Therefore, the total energy consumed by the microcontroller to switch from the transceiver off state to the data transfer mode is $0.86$\,ms$\times10.25$\,mA$\times V_{cc}$, i.e., $8.8\!\times\!V_{cc}$ microjoules.

Denote the overhead energy generated by the microcontroller between the transmission of packet $j$ and $j+1$ by $E^j_{ov}$. Based on the above analysis, $E^j_{ov}$ can be estimated as follows:
 
\begin{equation}
\label{eq:6B-01}
	E^j_{ov} = 0.2\,V^{end,\,j}_{cc}\,\frac{\left(C^j_c+4\,\text{mA}\right)}{2} + 8.8V^{end,\,j}_{cc},
\end{equation}
where $V^{end,\,j}_{cc}$ is the power supply voltage measured at the end of the $j^{th}$ packet transmission. The units of $E^j_{ov}$, $V^{end,\,j}_{cc}$, and $C^j_c$ in (\ref{eq:6B-01}) are $\mu$J, $V$, and mA, respectively.

In Appendix~\ref{subapp:OverEng}, we introduced how to calculate the overhead energy generated by the IEEE 802.15.4 protocol after considering the impact of ESC voltage reduction and $E_{ov}$ on system power consumption.


%% file: EngMod.tex
\section{Energy Consumption Model}
\label{sec:EngMod}

In this section, we utilize the results obtained in previous sections and propose a comprehensive energy consumption model for renewable radio energy powered IoT devices. In this model, the energy received by the device during data transmission will be ignored because the energy harvesting rate in an outdoor RF environment is at least two orders of magnitude lower than the system power consumption, as described in Section~\!\ref{sec:Env}.

Assume that an IoT device sends a total number of $N$ consecutive packets in each active transmission cycle. We show the model of cumulative energy consumption in Fig.~\!\ref{fig:EngConMod}. Denote the bit $i$ in the MSDU frame of the packet $j$ by $b_{(i, j)}$. Let $E_t(L^N_e, N)$ represent the accumulative energy consumption on sending data from $b_{(1, 1)}$ to $b_{(L^N_e, N)}$. The $E_t(L^N_e, N)$ for sending each packet $j$ is composed of two parts, namely, the energy overhead $E^j_{oe}$ and the energy consumed on sending the MSDU frame $E^j_c$. The calculation of $E^j_c$ can refer to (\ref{eq:appB-05}) in Appendix~\!\ref{subapp:OverEng}. The energy overhead varies for during consecutive packet transmissions. The calculation of $E^j_{oe}$ is described as follows:

\begin{figure}[htb]
\centerline{\includegraphics[width=7.7cm]{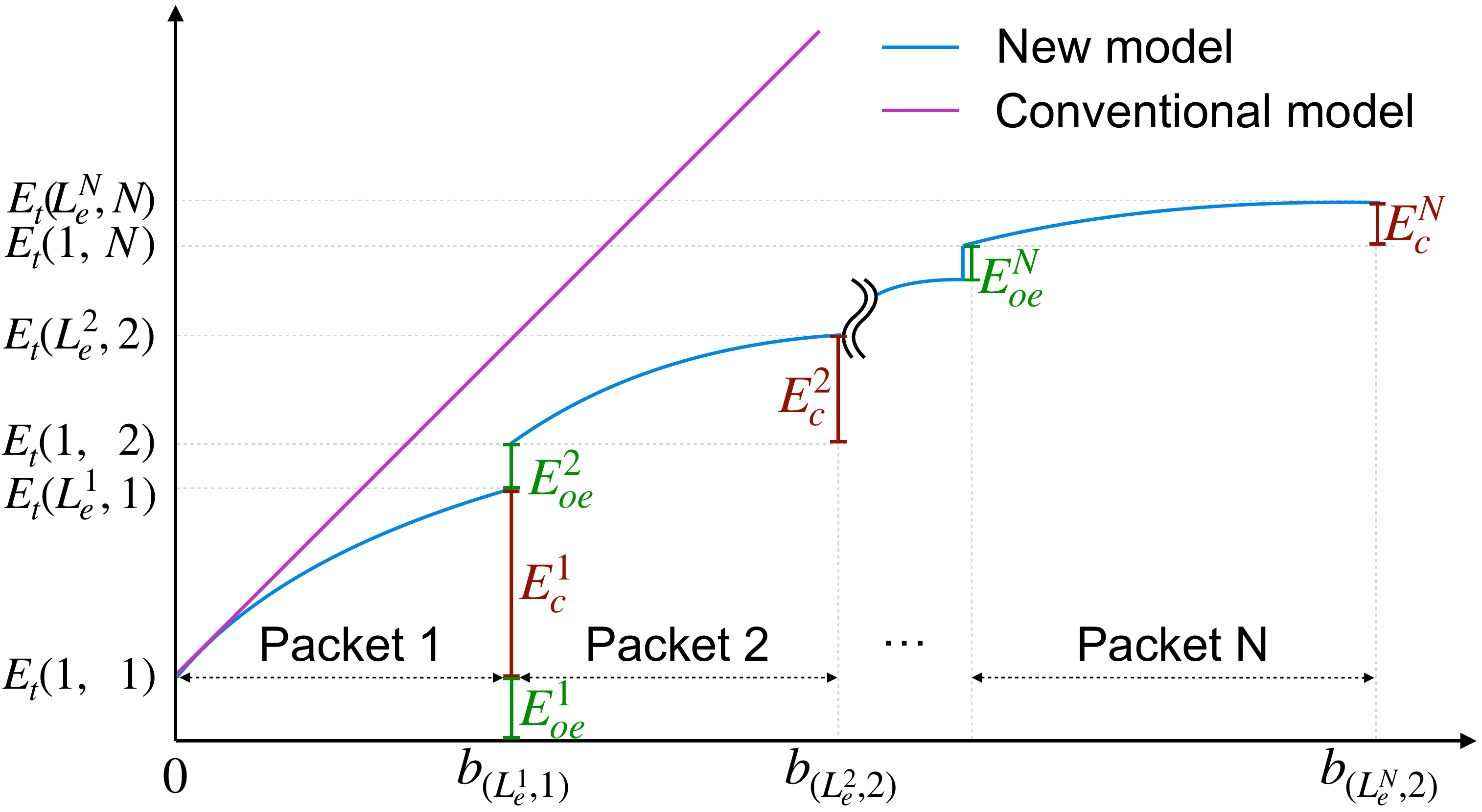}}
  \caption{Comparison between new and old energy consumption models.}
  \label{fig:EngConMod}
\end{figure}

\vspace{0.1cm}
\begin{adjustwidth}{-0.77cm}{0cm}
\begin{description}
\setlength{\labelsep}{-0.95em}
\itemsep 0.07cm
  \item[a)] $E^1_{oe}$: The energy overhead involved in sending the first packet. It consists of two parts: (a) the system wakeup-energy, and (b) the protocol overhead, which is denoted by $E^1_p$ in Appendix~\!\ref{subapp:OverEng}. The wake-up energy can be obtained by replacing $V_{cc}$ in (\ref{eq:5A_02}) to $V^{1,j}_{cc}$. How to calculate $E^1_p$ can be found in (\ref{eq:appB-09}) of Appendix~\!\ref{subapp:OverEng}.
\vspace{0.1cm}
  \item[b)] $E^j_{oe}$ for $j=2,3\ldots,N-1$: The energy overhead generated by the device when sending packet $2$ to $N-1$. It has of two parts: (a) the protocol overhead denoted by $E^j_p$, and (b) the overhead between transmitting packet $j-1$ and packet $j$, i.e., $E^j_{ov}$. How to calculate $E^j_p$ can refer to (\ref{eq:appB-09}) in Appendix~\!\ref{subapp:OverEng}. In addition, $E^j_{ov}$ is available in (\ref{eq:6B-01}), where $V^{end,\,j}_{cc}$ needs to be replaced by $V^{i_4,\,j}_{cc}$, which can be calculated by (\ref{eq:appB-08}) in Appendix~\!\ref{subapp:OverEng}.
\vspace{0.1cm}
  \item[c)] $E^N_{oe}$: The energy overhead generated by the device to transmit the last packet. It consists of two parts: (a) the protocol overhead denoted by $E^N_p$, and (b) the sleep energy generated by the system, i.e., $E_{ds}$.  How to calculate $E^N_p$ can refer to (\ref{eq:appB-09}) in Appendix~\!\ref{subapp:OverEng}. Additionally, $E_{ds}$ is available in (\ref{eq:5A_03}), where $V_{cc}$ and $C_c$ need to be replaced by $V^{i4,\,N}_{cc}$ and $C^{N}_c$, respectively; $V^{i4,\,N}_{cc}$ can be obtained via (\ref{eq:appB-08}) in Appendix~\!\ref{subapp:OverEng}.
\end{description}
\end{adjustwidth}
\vspace{0.1cm}

As illustrated in Fig.~\!\ref{fig:EngConMod}, the profile of $E_t$ has two important characteristics in the new energy consumption model. First, $ E_t $ is a discontinuous curve in an active cycle due to the energy overhead  generated before each data transmission. To be specific, whenever an IoT device sends a new packet, the system and communication protocol will involve additional overhead energy, resulting in a step increase in energy consumption. Second, each piece of $E_t$ does not increase linearly over time. As analyzed in (\ref{eq:6A-01}), the energy consumption of the IoT device decreases as the voltage of the ESC drops during continuous data transmission. Therefore, when sending a packet, the gradient of $E_t$ gradually decreases during data transmission. 

Due to the small capacity of the ESC and the low power density of renewable radio energy, the impact of the above two characteristics on power management cannot be ignored. In order to efficiently use the received energy, the IoT device needs to first determine the transmission power of each data packet based on the channel quality. Afterward, the device can then calculate the required supply current through the inverse function of (\ref{eq:5B-01}). Finally, the total energy consumed by the device when sending continuous packets can be accurately estimated based on the new model.

In Fig.~\!\ref{fig:EngConMod}, we also compare the new energy consumption model with the conventional one. In the conventional model, it is usually assumed that for a given transmission power and data rate, $E_t$ will increase linearly over time. As a result, the profile of accumulative energy consumption is a piecewise linear curve in an energy tunnel, as introduced in Section~\!\ref{sec:RelatedWork}. In the new model, $E_t$ becomes a nonlinear function of $b_{(i,\,j)}$. When the ESC capacity is low, the conventional model may overestimate the power consumption of IoT devices, making the power management strategy conservative. Furthermore, as revealed in the new model, after considering the energy overhead, a good transmission strategy should fill the MSDU frame as much as possible to improve energy efficiency, which is ignored in the conventional model.

%% file: Conclusion.tex
\section{Conclusions}
\label{sec:conclusion}
In this paper, we proposed a new model to describe the energy harvesting process and the power consumption of communication modules for renewable RF energy powered IoT devices. All parameters are based on experiment results to make the proposed model accurate.

As investigated in this paper, the intensity of renewable radio energy can have random and significant fluctuations in the outdoor environment. However, the ESC charging process can remain stable in such an environment, which makes the energy harvesting process highly predictable. To use the received energy efficiently, we used the Microchip ATmega256RFR2 microcontroller as an example to analyze the overhead energy generated by the hardware and communication protocol in the real world. Experimental results show that the supply voltage and MSDU payload can significantly affect the overhead energy, which has been carefully considered in the new model.

We believe that the model and results obtained in this paper can help researchers design more realistic and efficient power manage methods and data transmission strategies for renewable radio energy powered IoT devices.

%% file: Appendix.tex
\appendix
\section{Appendix}
\label{app:appendix}

\subsection {Energy Consumption on Successive Data Transmission}
\label{subapp:EngCon}
Assuming that the ESC voltage before transmitting bit $i$ is $V_{cc}^i$, then the energy consumed when sending the $i^{th}$ data bit is 
\begin{equation}
\label{eq:appA-00}
  E_s^i=V_{cc}^i\,C_c\,T_b. 
\end{equation}
The difference in energy consumption between two consecutive data bits is:
\begin{equation}
\label{eq:appA-01}
  E_s^{i-1}- E_s^i=\left(V_{cc}^{i-1}-V_{cc}^i\right) C_c \,T_b. 
\end{equation}
According to the energy storage equation of the capacitor, we also have that  
\begin{equation}
\label{eq:appA-02}
  \displaystyle\frac{1}{2}C\left[\left(V_{cc}^{i-1}\right)^2-\left(V_{cc}^i\right)^2\right]=E_s^i. 
\end{equation}
Based on (\ref{eq:appA-00}) and (\ref{eq:appA-02}), an iterative expression about $E^i_s$ can be obtained. Next, a method of simplifying the iterative expression (\ref{eq:appA-00}) and (\ref{eq:appA-02}) is provided to directly calculate $E^i_s$ for any $i\geq2$ based on $E^1_s$.

We substitute (\ref{eq:appA-01}) into (\ref{eq:appA-02}) and then perform a simple derivation to get that
\begin{equation}
\label{eq:appA-03}
  V_{cc}^{i-1}=\left[\left(V_{cc}^i\right)^2+\left(\displaystyle\frac{C_c T_b}{C}\right)^2\right]^{\frac{1}{2}}+\displaystyle\frac{C_c T_b}{C}. 
\end{equation}
In (\ref{eq:appA-03}), the product of $C_c$ (mA) and $T_b$ ($\mu$s) is in the order of $10^{-8}$ or lower, which is much smaller than the capacity (mF) of the supercapacitor used in a sustainable IoT device. In this case,  $(C_cT_b/C)^2 \approx 0$ and (\ref{eq:appA-02}) can be simplified as
\begin{equation}
\label{eq:appA-04}
  V_{cc}^{i-1}=V_{cc}^i+\displaystyle\frac{C_c T_b}{C}. 
\end{equation}
Substituting  (\ref{eq:appA-04}) into (\ref{eq:appA-01}), we have that 
\begin{equation}
\label{eq:appA-05}
  E_s^{i-1}- E_s^i=\displaystyle\frac{\left(C_c T_b\right)^2}{C}. 
\end{equation}

As can be observed from (\ref{eq:appA-05}), the difference between the energy consumption of transmitting consecutive data bits is constant. As a result, $\left\{E_s^1, E_s^2,\ldots, E_s^i\right\}$ forms an arithmetic progression. Accordingly, given a transmission power and a data rate, we can use the initial value of the energy consumption, $E^1_s$, and the common difference of the arithmetic progression, $(C_c T_b)^2/2$,  to directly calculate the energy consumed of sending the $i^{th}$ data bit as follows:
\begin{equation}
\label{eq:appA-06}
  E_s^i = E_s^1- \displaystyle\frac{(i-1)\left(C_c T_b\right)^2}{C}.
\end{equation}

\begin{figure}[htb]
\centerline{\includegraphics[width=6.5cm]{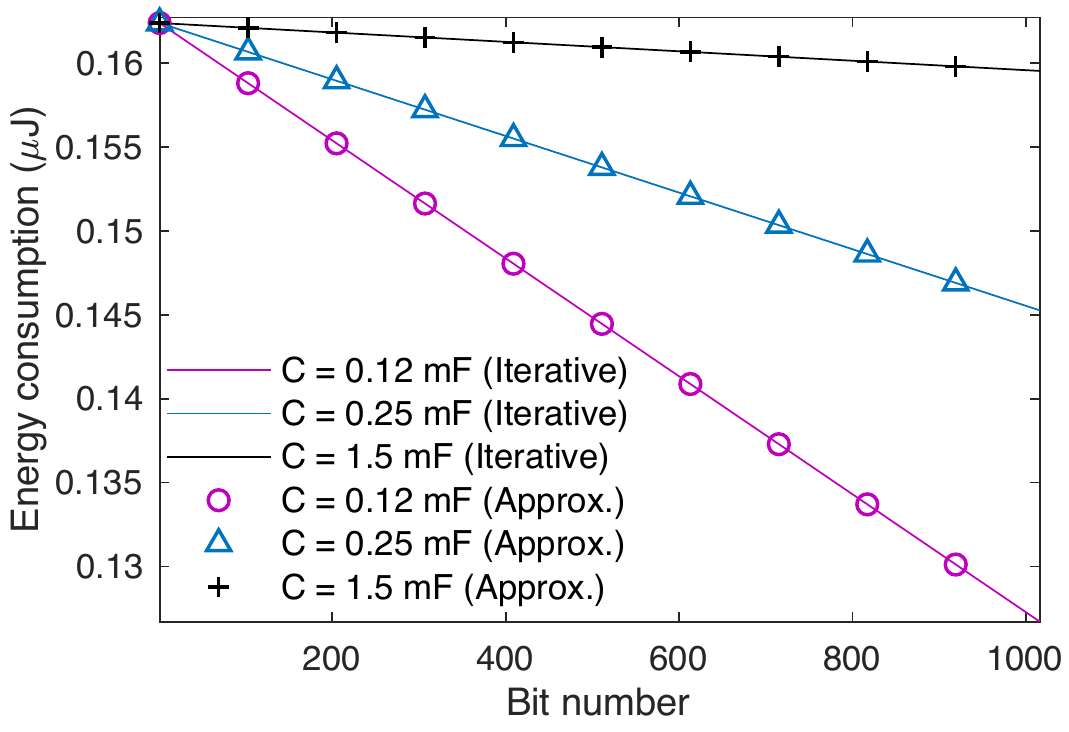}}
  \caption {Energy consumption on successive data transmission, where  $r_d=250$\,kbps, $C_c=16.24$\,mA, and $V^1_{cc}=2.5$\,V.}
  \label{fig:Pow_BitNum}
\end{figure}

In Fig.~\!\ref{fig:Pow_BitNum}, we compare the true energy consumption calculated by the iterative express (\ref{eq:appA-00}) and (\ref{eq:appA-02}) and the approximate value obtained from (\ref{eq:appA-06}). From the figure, it can be observed that the results  from the arithmetic expression matches the true values very well. Therefore, $E_s^i$ can be considered as a linear function of the bit number.

\subsection {Energy Overhead}
\label{subapp:OverEng}
Let $E^j_p$ represnt the overhead energy generated by the communication protocol when sending the packet $j$. It consists of three parts: the energy consumed by the IoT device to sent PHY preamble, MHR frame, and FCS frame, which are denoted by $E^j_{PHY}$, $E^j_{MHR}$, and $E^j_{FCS}$, respectively. Denote the MSDU payload of packet $j$ by $L^j_e$.  Next, we introduce how to calculate each part of $E^j_p$ based on (\ref{eq:6A-01}).

 \begin{figure}[htb]
\centerline{\includegraphics[width=7cm]{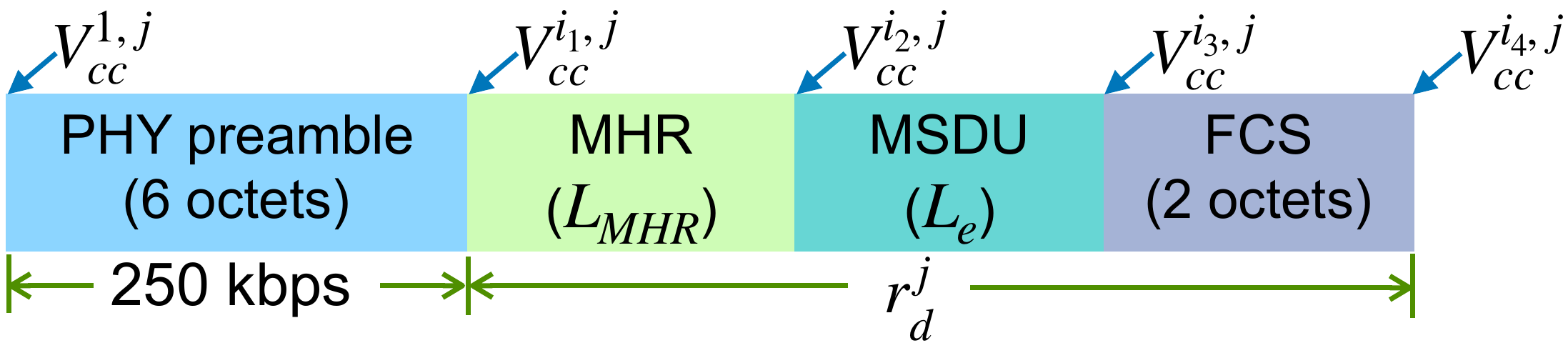}}
  \caption {The data structure of IEEE 802.15.4, where $V^{i,j}_{cc}$ is the ESC voltage before transmitting the bit $i$ in the packet $j$.}
  \label{fig:EngOverhead}
\end{figure}

According to the data structure of the IEEE 802.15.4 protocol, all overhead frames except FCS are sent before MSDU, as illustrated in Fig.~\!\ref{fig:EngOverhead}, where $i_1=48$, $i_2=i_1+L_{MHR}$, $i_3=i_2+L_{e}$, and $i_4= i_3+L_{FCS}$. In addition, the PHY preamble, including PHR and SHR, needs to be transmitted at $250$\,kbps of data rate. Based on (\ref{eq:6A-01}), it can be obtained that the energy consumed by the device to send the PHY preamble of packet $j$ is:

\begin{equation}
\label{eq:appB-01}
  E^j_{PHY}=\!\displaystyle\sum^{i_1}_{i=1}E^{i,j}_s\!=\! \displaystyle\frac{i_1V^{1,j}_{cc}C^j_c}{r^j_d}+\displaystyle\frac{i_1\,(i_1\!-\!1)}{2\,C^2}\!\!\left(\displaystyle\frac{C^j_c}{\text{250 kbps}}\!\right)^4.
\end{equation}
In Fig.~\!\ref{fig:EngOverhead}, based on the energy storage equation of the capacitor, we can calculate $V^{j,i_1}_{cc}$, which is also the ESC voltage after transmitting PHY preamble, as follows:
\begin{equation}
\label{eq:appB-02}
  V^{i_1,j}_{cc}=\sqrt{\left(V^{1,j}_{cc}\right)^2-\frac{2E^j_{PHY}}{C}}.
\end{equation}

Similar to (\ref{eq:appB-01}), the energy consumed by the device to transmit MHR frame is:
\begin{equation}
\label{eq:appB-03}
\begin{array}{lll}
\vspace{0.17cm}	
	E^j_{MHR} \!\!\!\!\!\!&=&\!\!\!\!\!\! \displaystyle\sum^{i_2}_{i=i_1+1}\!\!\!\!E^{i,j}_s\\
	\!\!\!&=&\!\!\! \displaystyle\frac{L_{MHR}V^{i_1,j}_{cc}C^j_c}{r^j_d}\!+\!\displaystyle\frac{L_{MHR}(L_{MHR}\!-\!1)}{2\,C^2}\!\!\left(\displaystyle\frac{C^j_c}{r^j_d}\!\right)^4.
\end{array}
\end{equation}
Afterward, using the energy storage equation again, we have that 
\begin{equation}
\label{eq:appB-04}
  V^{i_2,j}_{cc}=\sqrt{\left(V^{i_1,j}_{cc}\right)^2-\frac{2E^j_{MHR}}{C}}.
\end{equation}
Let $E^j_c$ represent the total energy consumed by the device to send the MSDU frame (i.e., the effective data), which can be calculated as follows:
\begin{equation}
\label{eq:appB-05}
	E^j_c = \displaystyle\sum^{i_3}_{i=i_2+1}\!\!\!\!E^{i,j}_s = \displaystyle\frac{L^j_eV^{i_2,j}_{cc}C^j_c}{r^j_d}\!+\!\displaystyle\frac{L^j_e(L^j_e\!\!-\!\!1)}{2\,C^2}\!\!\left(\displaystyle\frac{C^j_c}{r^j_d}\!\right)^4.
\end{equation}
Similar to (\ref{eq:appB-04}), we have that :
\begin{equation}
\label{eq:appB-06}
  V^{i_3,j}_{cc}=\sqrt{\left(V^{i_2,j}_{cc}\right)^2-\frac{2E^j_e}{C}}.
\end{equation}
Thereafter, the energy consumption for transmitting FCS frames is:
\begin{equation}
\label{eq:appB-07}
	E^j_{FCS} = \displaystyle\sum^{i_3+3}_{i=i_3+1}\!\!\!\!E^{i,j}_s = \displaystyle\frac{16\,V^{i_2,j}_{cc}\,C^j_c}{r^j_d}\!+\!\displaystyle\frac{16\times15}{2\,C^2}\!\!\left(\displaystyle\frac{C^j_c}{r^j_d}\!\right)^4.
\end{equation}
Similar to (\ref{eq:appB-06}), after sending the FCS frame, the ESC voltage becomes:
\begin{equation}
\label{eq:appB-08}
  V^{i_4,j}_{cc}=\sqrt{\left(V^{i_3,j}_{cc}\right)^2-\frac{2E^j_{FCS}}{C}}.
\end{equation}
Eventually, the overhead energy generated by the communication protocol to send the packet $j$ is
\begin{equation}
\label{eq:appB-09}
  E^j_p = E^j_{PHY}+E^j_{MHR}+E^j_{FCS},
\end{equation}
where $E^j_{PHY}$, $E^j_{MHR}$, and $E^j_{FCS}$ are calculated in (\ref{eq:appB-01}), (\ref{eq:appB-03}), and (\ref{eq:appB-07}), respectively.